# Evaluation of bulk-interface contributions to Edelstein magnetoresistance at metal/oxide interfaces

Junyeon Kim[1], Yan-Ting Chen[1], Shutaro Karube[1,2], Saburo Takahashi[3], Kouta Kondou[1], Gen Tatara[1], and YoshiChika Otani[1,2]

[1]*Center for Emergent Matter Science, RIKEN, Wako, Saitama 351-0198, Japan*

[2]*Institute for Solid State Physics, University of Tokyo, Kashiwa, Chiba 277-8581, Japan*

[3]*Institute for Materials Research, Tohoku University, Sendai, Miyagi 980-8577, Japan*

We report a systematic study on Edelstein magnetoresistance (Edelstein MR) in $Co_{25}Fe_{75}/Cu/Bi_2O_3$ heterostructures with a strong spin-orbit interaction at the $Cu/Bi_2O_3$ interface. We succeed in observing a significant dependence of the Edelstein MR on both Cu layer thickness and temperature, and also develop a general analytical model considering distinct bulk and interface contributions on spin relaxation. Our analysis, based on the above model, quantitatively illustrates a unique property of the spin transport near the Rashba interface, revealing a prominent role of the spin relaxation process by determining the ratios of the spin relaxation inside and outside the interface. We further find the characteristic spin transport is unaffected by temperature. Our results provide an essential tool for exploring the transport in a system with spin-momentum-locked two-dimensional states.

Spin-orbit interactions (SOIs) have brought about a paradigm shift in magnetization manipulation, where the spin current can be generated by the spin Hall effect (SHE) in the bulk (semiconductor, and heavymetal) [1] and the Edelstein effect at the spin-momentum-locked states as shown in Fig. 1(a) at a Rashba interface and a topological insulator (TI) surface [2-4]. Recent studies have demonstrated that the charge-to-spin current conversion originating from the SHE and/or the Edelstein effect has enabled efficient magnetization manipulation via spin-orbit torques in various heterostructures [5-14]. Also, the SOI leads to a spin-to-charge current conversion as a form of the inverse SHE [15] and/or the inverse Edelstein effect [Fig. 1(b)] [16-18]. Interestingly, the spin/charge current interconversion due to the SOI also gives rise to a modulation of the longitudinal resistance, e.g., spin Hall magnetoresistance (SMR) [19] and Edelstein magnetoresistance (Edelstein MR) [20]. These novel magnetoresistances (MRs) are manifested in various bilayers such as spin Hall materials or Rashba interfaces neighboring a magnetic insulator [21, 22] or a ferromagnetic metal (FM) [23-27]. The novel MRs originate through a spin/charge current interconversion combined with anisotropic spin absorption to the FM depending on its magnetization direction, as depicted in Fig. 1(c). Particularly, the SMR enables us to evaluate the conversion efficiency, as well as to clarify the underlying properties on the spin transport [23]. The Edelstein MR is also expected to be quite helpful in understanding the underlying physics, although we do not yet have a general model.

Despite the many similarities, there is a significant difference between the SHE and the Edelstein effect. In bulk spin Hall materials, the spin-to-charge current conversion occurs as an exactly reversed process of the charge-to-spin current conversion satisfying the Onsager reciprocity. Experimental studies also supports the equivalence of the conversion efficiency between the direct and inverse SHEs [28]. However, the spin/charge current interconversion originating from the Edelstein effect is much more complex. Recent experimental studies have shown that the charge-to-spin current conversion is more than 100 times more efficient than the spin-to-charge current conversion in a $Bi_2Se_3$ topological surface [29, 30]. Moreover, a discrepancy in the conversion efficiency is broadly observed in several TI systems [13, 14, 17, 31, 32]. Although there have been many discussions regarding a solution for the discrepancy, it is still been an unsolved question.

Recently, Zhang *et al.* have provided a theoretical framework to understand the spin/charge current interconversion originating from the Edelstein effect by taking into account the spin relaxation processes occurring simultaneously in the bulk and at the interface [33]. As depicted in Figs. 1(a) and 1(b), both direct and inverse Edelstein effects (DEE and IEE) accompany a shift in the Fermi circle (spin accumulation). At this moment, the adjacent bulk (or the interface) contribution to the relaxation of accumulated spins is a crucial factor in determining the conversion efficiency of the DEE (or the IEE). Considering that the bulk and the interface contributions to the spin relaxation are independent, the role of spin relaxation on the spin/charge current interconversion implies that the Onsager reciprocity does not hold for the DEE and the IEE.

In this Rapid Communication, we study systematically the influence of spin relaxation on the Edelstein MR in $Co_{25}Fe_{75}/Cu/Bi_2O_3$ heterostructures with a $Cu/Bi_2O_3$ Rashba interface, and develop a general model for the Edelstein MR by considering the spin relaxation both at the Rashba interface and the adjacent bulk. In the model, we used the ratio of the spin relaxation inside to the outside of the Rashba interface. This analysis enables us to understand how the accumulated spins diffuse out of the interface, and brings a general framework of the quantitative description of the characteristic spin relaxation near the spin-momentum-locked state.

For this purpose, we fabricated Hall bar and waveguide devices [Figs. 2(a) and (b)] on a Si substrate by means of photolithography combined with e-beam evaporation. The typical dimensions of the Hall bar devices are 10 μm in width and 800 μm separation between the Hall crosses. The devices consist of $Co_{25}Fe_{75}$ (5 nm)/Cu (0-30 nm)/$Bi_2O_3$ (20 nm) multilayers e-beam evaporated on a Si substrate. Both the Hall bar and the waveguide devices are deposited simultaneously. Note that the easy axis of the $Co_{25}Fe_{75}$ layer is in plane regardless of the Cu thickness.

Our recent study has shown that an efficient spin-to-charge current conversion can be produced by the IEE at the Cu/$Bi_2O_3$ interface [34]. To verify the presence of IEE in the present structure, we carried out the spin pumping measurement using the waveguide devices [15, 35, 36]. We then evaluated the effective Rashba parameter $\alpha_{R\_eff}$ for the devices with various Cu thicknesses $d_{Cu}$ as shown in Fig. 2(c) [37-39]. Although $|\alpha_{R\_eff}|$ depends on $d_{Cu}$ in a thin regime, eventually $|\alpha_{R\_eff}|$ reaches ~0.4 eVÅ when $d_{Cu}$ becomes sufficiently thick. We think $|\alpha_{R\_eff}|$ dependence on $d_{Cu}$ in a thin regime comes from the improvement of crystallization of the Cu as $d_{Cu}$ increases. The value of saturated $|\alpha_{R\_eff}|$ in our devices is comparable to the reported values for a Py/Cu/$Bi_2O_3$ heterostructure [34].

We measured the longitudinal MR of the Hall bar devices by means of a conventional four-terminal method using a physical property measurement system (PPMS). The applied external magnetic field was rotated as a function of either angle $\alpha_{xy}$ in the $x$-$y$, $\beta_{yz}$ in the $y$-$z$, or $\gamma_{xz}$ in the $x$-$z$ plane as shown in Fig. 3(a).

The corresponding modulation of the resistance by the external field can be expressed as

$$R = R_0 + \Delta R_{AMR} m_x^2 - (\Delta R_{EdMR} + \Delta R_{etc}) m_y^2, \tag{1}$$

where $R_0$ is the intrinsic resistance, and $m_{x(y)}$ are the Cartesian components of the magnetization vector of the FM [19, 23]. $\Delta R_{AMR}$ and $\Delta R_{EdMR}$ represent the resistance contributions of the anisotropic MR (AMR) and the Edelstein MR, respectively. $\Delta R_{etc}$ represents an additional magnetoresistance as shown in the CoFe/Cu bilayers [37, 40], possibly attributable to recently found MRs such as the hybrid MR [41], the interfacial AMR [42, 43], or a mixture of them. Both of them are known to be irrelevant in the spin/charge current interconversion in the system. Additionally, the geometrical size effect (GSE), the difference in resistance measured under the magnetic field along the $y$ and $z$ directions in the FM layer [44], also has an influence on $\Delta R_{etc}$. Except for a very thin Cu thickness regime, we assume $\Delta R_{etc} \sim 0.01$ % in the CoFe/Cu/Bi$_2$O$_3$ structures is independent of $d_{Cu}$ considering the MR in CoFe/Cu bilayers.

The resistances as a function of the rotation angles, $\Delta R(\alpha_{xy})$, $\Delta R(\beta_{yz})$, and $\Delta R(\gamma_{xz})$ are shown in Figs. 3(b) and 3(c). The applied field is set at 60 kOe which is large enough to uniformly saturate the magnetization of the FM layer. The resistance shows a sinusoidal dependence as a function of angles in accordance with Eq. (1) [Figs. 3(b) and 3(c)]. $\Delta R(\beta_{yz})$ for $d_{Cu} = 0$ nm takes a minimum when the field is along the $z$-axis ($\beta_{yz} = 0°$ or 180°) due to the geometrical size effect [Fig. 3(b)]. Following the enlargement $|\alpha_{R\_eff}|$ with increasing $d_{Cu}$, the Edelstein MR overcomes the geometrical size effect, showing a minimum $\Delta R(\beta_{yz})$ at the field along $y$ axis ($\beta_{yz} = 90°$ or 270°) for $d_{Cu}=4.7$ nm. [Fig. 3(c)].

The Edelstein MR ratio $\Delta R_{EdMR}/R$ divided by the total resistance $R$ is plotted as a function of $d_{Cu}$ [Fig. 4(b)]. The obtained $\Delta R_{EdMR}/R$ exhibits a peak when $d_{Cu} \sim 6$ nm, then it decreases with $d_{Cu}$. Note that the negative $\Delta R_{EdMR}/R$ in very thin $d_{Cu}$ comes from the dominant role of the GSE as shown in Fig. 3(b). To clarify this trend in $\Delta R_{EdMR}/R$, we developed a general analytical model as described below.

In order to build an analytical model, we formulate the Edelstein MR by considering with mechanism depicted in Fig. 1(c) [37]. As a first, we consider the charge-to-spin conversion via the DEE at the Cu/Bi$_2$O$_3$ interface. The generated spin accumulation by the DEE is diffused to the Cu layer, and finally it is absorbed by the CoFe layer anisotropically dependent of its magnetization orientation governed by the spin transfer torque (STT). Considering the boundary conditions at the CoFe/Cu and the Cu/Bi$_2$O$_3$ interfaces, we extract the spin accumulation and the net spin current inside the Cu layer by using the spin-diffusion equation [19]. Eventually, we obtain the converted extra charge current from the net spin current via the IEE, and formulate the modulation of the longitudinal resistance. We also consider the effect of current shunting, i.e., a part of the current flows in the CoFe layer, which makes no contribution to the spin/charge current interconversion. Most importantly, we take into account a characteristic property of the spin/charge current interconversion near the Rashba interface contributed by the spin relaxation as follows.

Accumulated spins at the Rashba interface diffuse either inside or outside the Rashba interface with different spin relaxation times as depicted in Fig. 4(a) [33]. The spin-flipping inside the Rashba interface is characterized by the spin flipping time $\tau_i$. On the other hand, the spin relaxation outside the Rashba interface is governed by the characteristic time $\tau_b$. Therefore, a total relaxation time of the nonequilibrium spin state at the Rashba interface $\tau_t$ is given by a relation $1/\tau_t = 1/\tau_i + 1/\tau_b$. Considering the fact that the inverse of the characteristic time corresponds to a probability, the accumulated spins undergo spin relaxation either at the Rashba interface or outside with a certain ratio. Here, we define the ratio $\eta$ of the spin relaxation at the outside of the Rashba interface to the total relaxation by $\eta \equiv (1/\tau_b)/(1/\tau_t)$. Similarly, the ratio for the spin-flip relaxation inside the Rashba interface is given by $1 - \eta = (1/\tau_i)/(1/\tau_t)$.

Based on the above-mentioned arguments, we developed the general analytical model for the Edelstein MR [37],

$$\frac{\Delta R_{EdMR}}{R} = \xi^2 \gamma_{EE} \lambda_{IEE} \frac{1}{et_I} \frac{G_r \text{sech}^2(d_{Cu}/l_{Cu})}{1 + 2\rho_{Cu} l_{Cu} G_r \tanh(d_{Cu}/l_{Cu})} \quad \text{with}$$

$$\xi \equiv \frac{1}{1 + \frac{\rho_{Cu} t_{CF}}{\rho_{CF} d_{Cu}}} \frac{1}{1 + \frac{\rho_I d_{Cu}}{\rho_{Cu} t_I}}, \tag{2}$$

$$\gamma_{EE} \equiv \eta \frac{3 m_e^* \alpha_{R\_eff}}{2e\hbar \varepsilon_F} \frac{1}{D(\varepsilon_F)}, \text{ and}$$

$$\lambda_{IEE} \equiv \frac{1}{1-\eta} \frac{\alpha_{R\_eff} \tau_t}{\hbar} = \frac{\alpha_{R\_eff} \tau_i}{\hbar},$$

where $\xi$ represents the current shunting, and $\gamma_{EE}$ and $\lambda_{IEE}$ represent the coefficients characterizing the DEE and the IEE, respectively. Note that $\gamma_{EE}$ accounts for the ratio $\eta$ of relaxed spins outside to the relaxed total spins. The density of states $D(\varepsilon_F)$ is a conversion factor between the number density of non-equilibrium spins and the spin accumulation [45]. Also $G_r$ is the real part of the spin mixing conductance between the FM and the Cu layer, and $e$ is the elementary charge, $m_e^*$ is the effective mass of electron at the Rashba interface, and $\varepsilon_F$ is the Fermi energy of the Cu [3]. $d_{Cu}$, $l_{Cu}$, and $\rho_{Cu}$ are the layer thickness, the spin diffusion length and the resistivity of the Cu layer, respectively, and $t_{CF(I)}$, and $\rho_{CF(I)}$ are the layer thickness and the resistivity of the CoFe layer (Cu/Bi$_2$O$_3$ interface). We used a typical interfacial thickness of $t_I$=0.4 nm in this model [16]. The values of $\rho_{Cu}$ in Eq. (2) are experimentally determined as shown in the inset of Fig. 4(b), displaying $\rho_{Cu}$ as a function of Cu thickness $d_{Cu}$. The significant $d_{Cu}$ dependence of $\rho_{Cu}$ attributed to additional scattering at the interface can be described by

$$\rho_{Cu} = \rho_\infty \left(1 + \frac{3}{8(d_{Cu} - h)}[l_\infty(1-p)]\right), \tag{3}$$

where $\rho_\infty$ and $l_\infty$ are the resistivity and mean free path for an intrinsic bulk Cu, respectively [22, 42]. The variables $h$ and $p$ represent the surface roughness in the unit of nanometer and the interfacial scattering rate which varies from 0 to 1, respectively.

With this model, we analyzed the $d_{Cu}$ dependence of $\Delta R_{EdMR}/R$, using with $|\alpha_{R\_eff}|$ from the spin pumping measurement, and $G_r = 3 \times 10^{15}$ $\Omega^{-1}$m$^{-2}$ for spin transport at the Co$_{25}$Fe$_{75}$/Cu from the literature [46]. $\Delta R_{EdMR}/R$ is well fitted to Eq. (2) by using the parameters $\tau_i$ and $\rho_I$ independent of $d_{Cu}$ [47]. Here it should be noted that we assume that $\tau_i$ is equal to the momentum relaxation time of 8.5 fs calculated from $\rho_\infty$ since the scattering inside the Rashba interface is strongly influenced by hybridization with the intrinsic bulk metallic state [17, 33, 48]. Similarly, $\rho_I$ is also assumed to be proportional to $\rho_\infty$. So far, several reports have commonly asserted that the effective mass of the conduction electrons at Rashba interfaces adjacent to the metal layer diminishes to several of tens percent of the electron mass [49-51]. In particular, the effective mass at the Cu/Bi interface, which has a similar property to the Cu/Bi$_2$O$_3$ interface, is reported to be 35 % of the electron mass [50]. Since the resistivity is proportional to the effective mass, likely $\rho_I$ is also reduced to 35 % of $\rho_\infty$. A fitting curve shown in Fig. 4(b) was obtained with $\rho_I/\rho_\infty=0.35$. As a result of the fitting, we obtained $\eta=33.9$ %. It means that only a third of the accumulated spins at the Cu/Bi$_2$O$_3$ interface moves to outside the Rashba interface. The corresponding characteristic times $\tau_b$ and $\tau_t$ are 16.6, and 5.6 fs, respectively.

We also carried out measurements on the temperature $T$ dependence of $\Delta R_{EdMR}/R$ in the $T$ range from 10 to 300 K. Figure 5(a) shows the $d_{Cu}$ dependence of $\rho_{Cu}$ at several temperatures. The resistivity $\rho_{Cu}$ was fitted to Eq. (3) by following the same method as for the inset of Fig. 4(b). The inset of Fig. 5(a) shows a typical metallic $T$ dependence of $\rho_\infty$. In contrast, $\Delta R_{EdMR}/R$ increases by 80 % at 10 K compared with the value at 300 K [Fig. 5(b)]. Note that we have normalized $\Delta R_{EdMR}/R$ for various temperatures by the value of $\Delta R_{EdMR}/R$ at room temperature. As shown in Fig. 5(c), $\Delta R_{EdMR}/R$ as a function of $d_{Cu}$ for different temperatures is fitted to Eq. (2) by assuming $T$-independent $\alpha_{R\_eff}$ and $G_r$ [52]. The negligible $T$ dependence of $\alpha_{R\_eff}$ is supported by an estimation based on the relation between $\alpha_{R\_eff}$, $\lambda_{IEE}$, and $\tau_i$ in Eq. (2). Here we considered the change of $\tau_i$ ($\propto 1/\rho_\infty$) and $\lambda_{IEE}$ as a function of $T$ from the inset of Fig. 5(a) and the literature [53], respectively. Interestingly, the obtained $\eta$ from the fitting for various values of $\rho_I/\rho_\infty$ [Fig. 5(d)] shows a negligible $T$ dependence irrespective of the values of $\rho_I/\rho_\infty$. It means the ratio of the spin relaxation at the outside of the Rashba interface is not so much affected by $T$ in this structure. Likely, the spin-flip process inside the interface ($\tau_i$) and the spin transition from the interface to the bulk ($\tau_b$) becomes less frequent in low $T$ by showing almost the same amount of change.

In summary, we studied the Cu layer thickness and temperature dependence of the Edelstein MR at the $Cu/Bi_2O_3$ interface by developing a general model. From the analysis of the Edelstein MR, we revealed the characteristic nature of spin transport at the Rashba interface to be governed by spin relaxation. The spin relaxation times are crucial parameters to determine the ratio of the spin relaxation inside and outside the Rashba interface. The ratio of the spin relaxation provides a different relation between the DEE and the IEE instead of the Onsager reciprocity.

The authors acknowledge H. Isshiki for various discussions. This work was supported by a Grant-in-Aid for Scientific Research on the Innovative Area, 'Nano Spin Conversion Science' (Grant No. 26103002).

**Figure captions**

FIG. 1. (a),(b) Shift of the Fermi contours when (a) the charge current or (b) the spin current is induced at the Rashba interface. (c) Mechanism of the Edelstein MR. Small (large) net spin current, large (small) generated extra charge current, and low (high) resistance is exhibited when $\vec{M} /\!/ \vec{\sigma}$ ($\vec{M} \perp \vec{\sigma}$), as shown in the left (right) panel, where $\vec{M}$ and $\vec{\sigma}$ represent the magnetization directions of the FM and the spin accumulation vector, respectively. Note that red and blue arrows with $e$- and $\Delta e$- represent the induced and generated flow of the charge electrons, respectively.

FIG. 2. (a), (b) Schematic illustrations of (a) the Hall bar and (b) the waveguide devices. As shown in (b), we lead the ferromagnetic resonance by applying an external field ($H_{ext}$) and radio-frequency Oersted field ($H_{rf}$) from the signal generator (S. G.) during the spin pumping measurement. (c) Absolute value of the obtained $\alpha_{R\_eff}$ plotted as a function of $d_{Cu}$. The dashed line is a guide that indicates the saturated values of $|\alpha_{R\_eff}|$.

FIG. 3. (a) Definition of the Cartesian axes and the rotating angles of the external magnetic field. The rotating angles on $x$-$y$, $y$-$z$ and $x$-$z$ planes are denoted by $\alpha_{xy}$, $\beta_{yz}$, and $\gamma_{xz}$, respectively. Here $\alpha_{xy}$, $\beta_{yz}$, and $\gamma_{xz}=0$ correspond to the field along to the $y$, $z$, and $z$ axis, respectively. (b),(c) Resistance with changing the angles of the field in devices with (b) $d_{Cu}$ =0 nm and (d) $d_{Cu}$=4.7 nm, respectively. Symbols with black squares, red circles, and blue triangle correspond to $\Delta R(\alpha_{xy})$, $\Delta R(\beta_{yz})$, and $\Delta R(\gamma_{xz})$, respectively.

FIG 4. (a) Red and black arrows represent the spin relaxation outside and inside the Rashba

interface. The former and the latter are governed by characteristic times $\tau_b$, and $\tau_i$, respectively. (b) $\Delta R_{EdMR}/R$ from room temperature plotted as a function of $d_{Cu}$. The solid line is a fitting curve with Eq. (2). Here the curve is obtained with $t_I$=0.4 nm and $\rho_I/\rho_\infty$=0.35. The inset shows $\rho_{Cu}$ plotted as a function of $d_{Cu}$. The solid line is a fitting curve by using Eq. (3).

FIG. 5. (a) $\rho_{Cu}$ plotted as a function of $d_{Cu}$ from several temperatures. Note that symbol colors and shapes are the same as those in (c). Solid lines are fittings made by using Eq. (3). The inset presents obtained $\rho_\infty$ as a function of $T$. (b) Typical normalized $\Delta R_{EdMR}/R$ plotted as a function of $T$ from a device with $d_{Cu}$=6.8 nm. The dashed line corresponds to 1. (c) $\Delta R_{EdMR}/R$ plotted as a function of $d_{Cu}$ for several temperatures. Solid lines represent fitting curves by using Eq. (2). (d) $\eta$ plotted as a function of $T$ for $\rho_I/\rho_\infty$=0.25, 0.35, and 0.45, respectively.

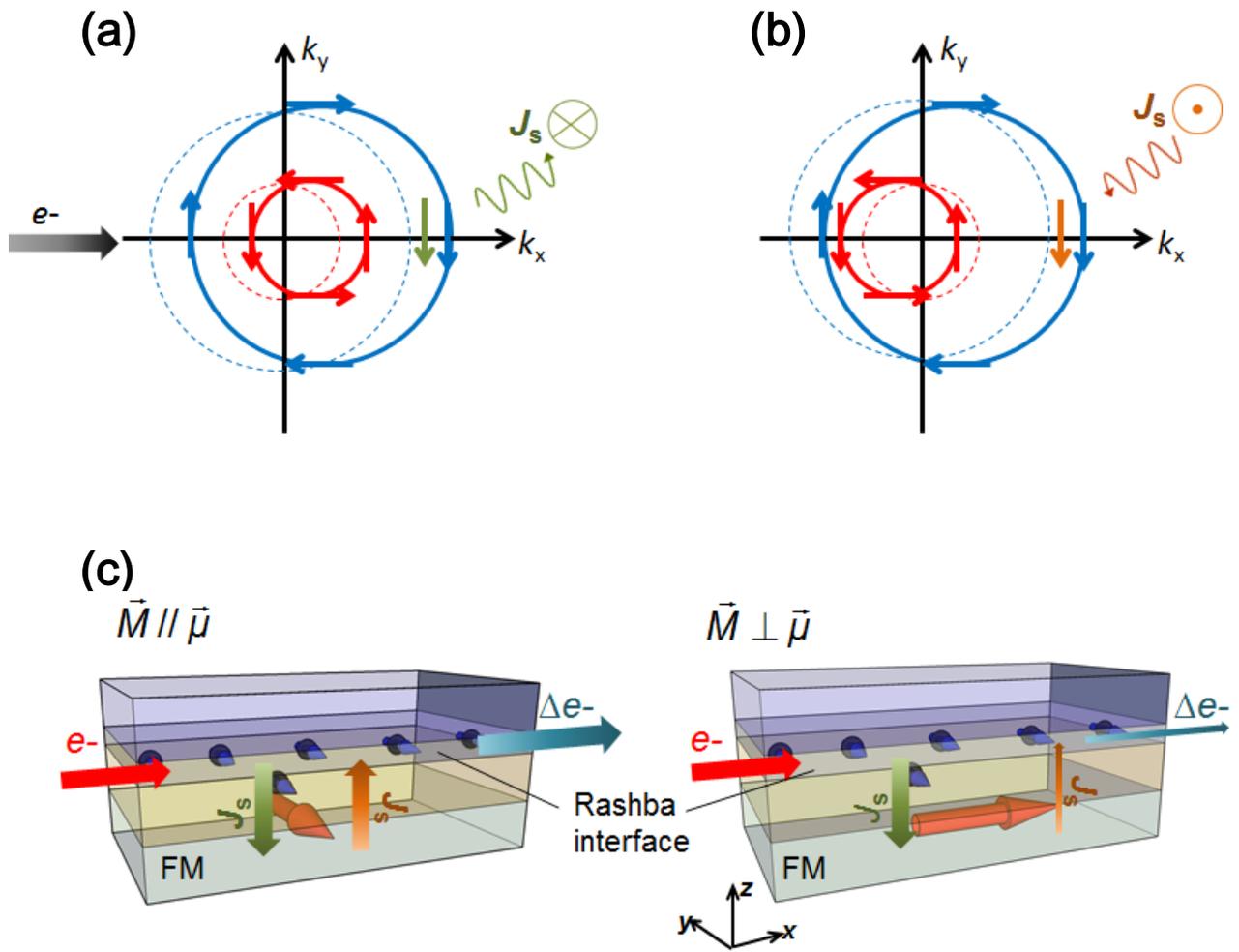

Fig. 1

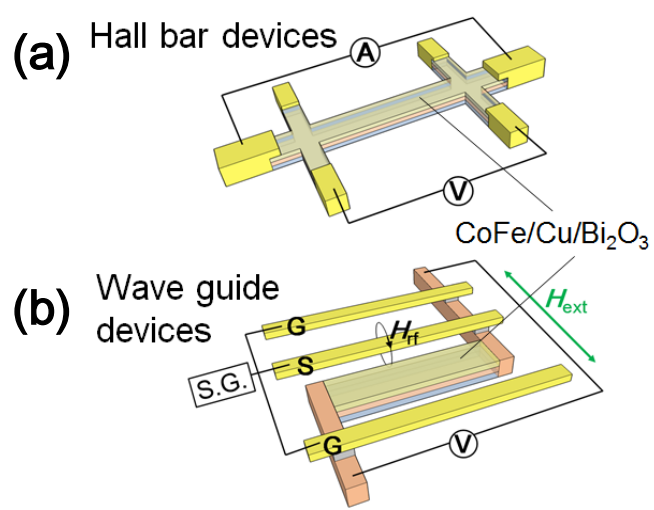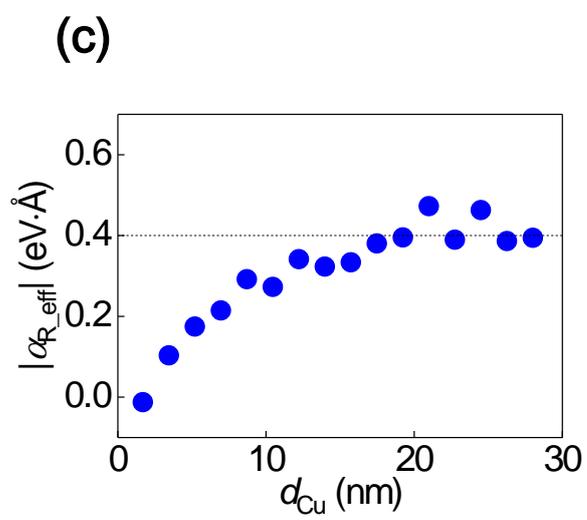

Fig. 2

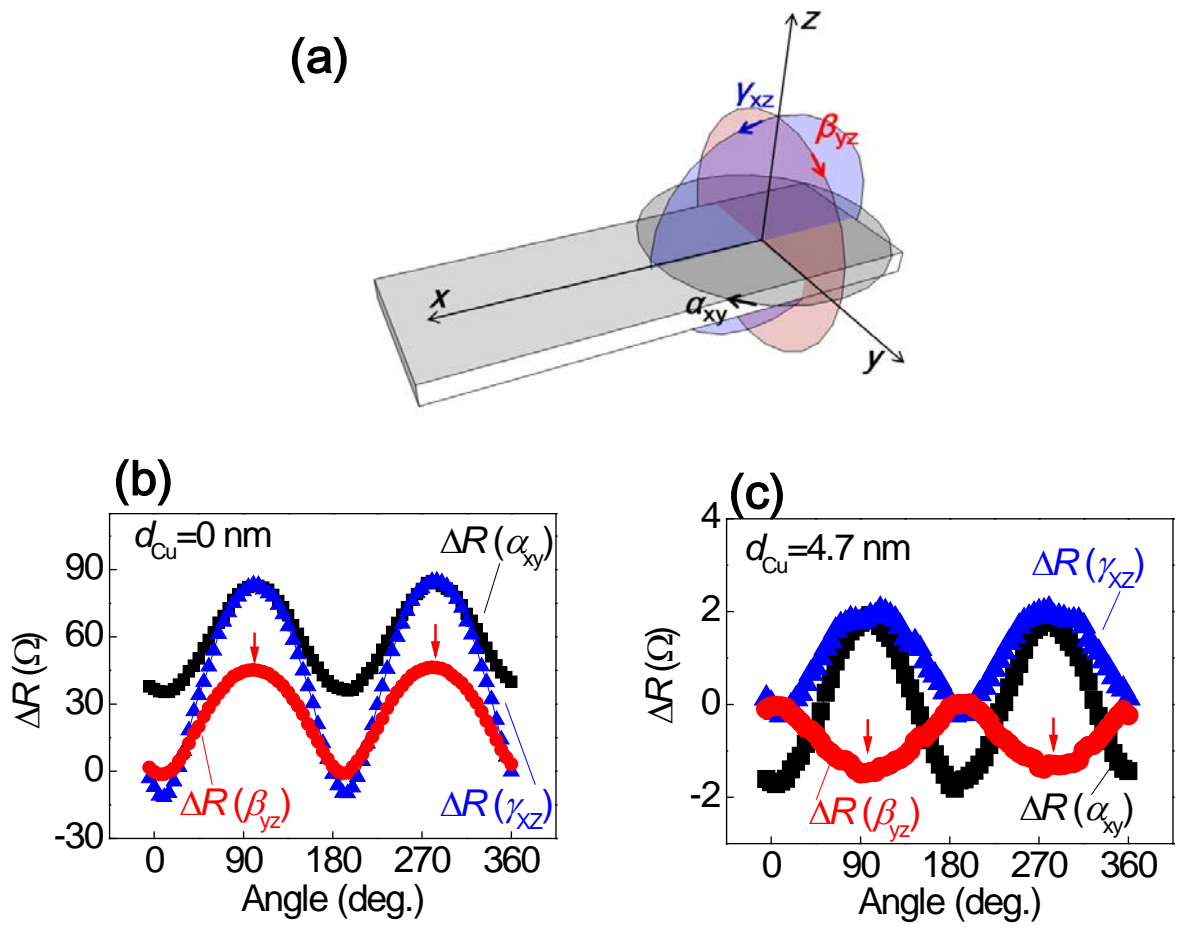

Fig. 3

(a) 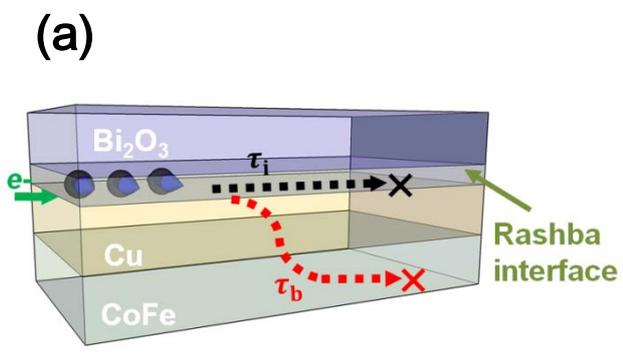

(b) 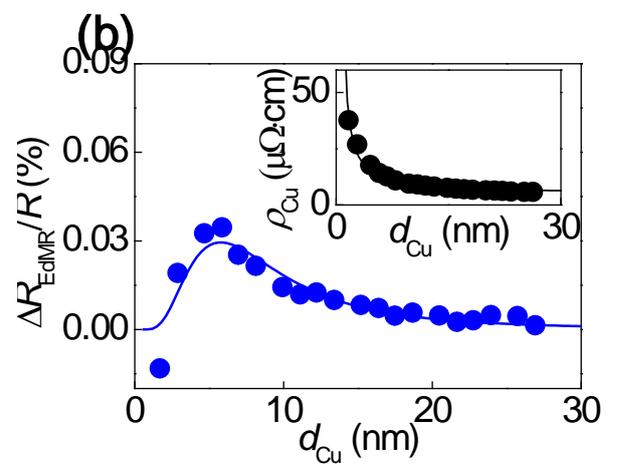

Fig. 4

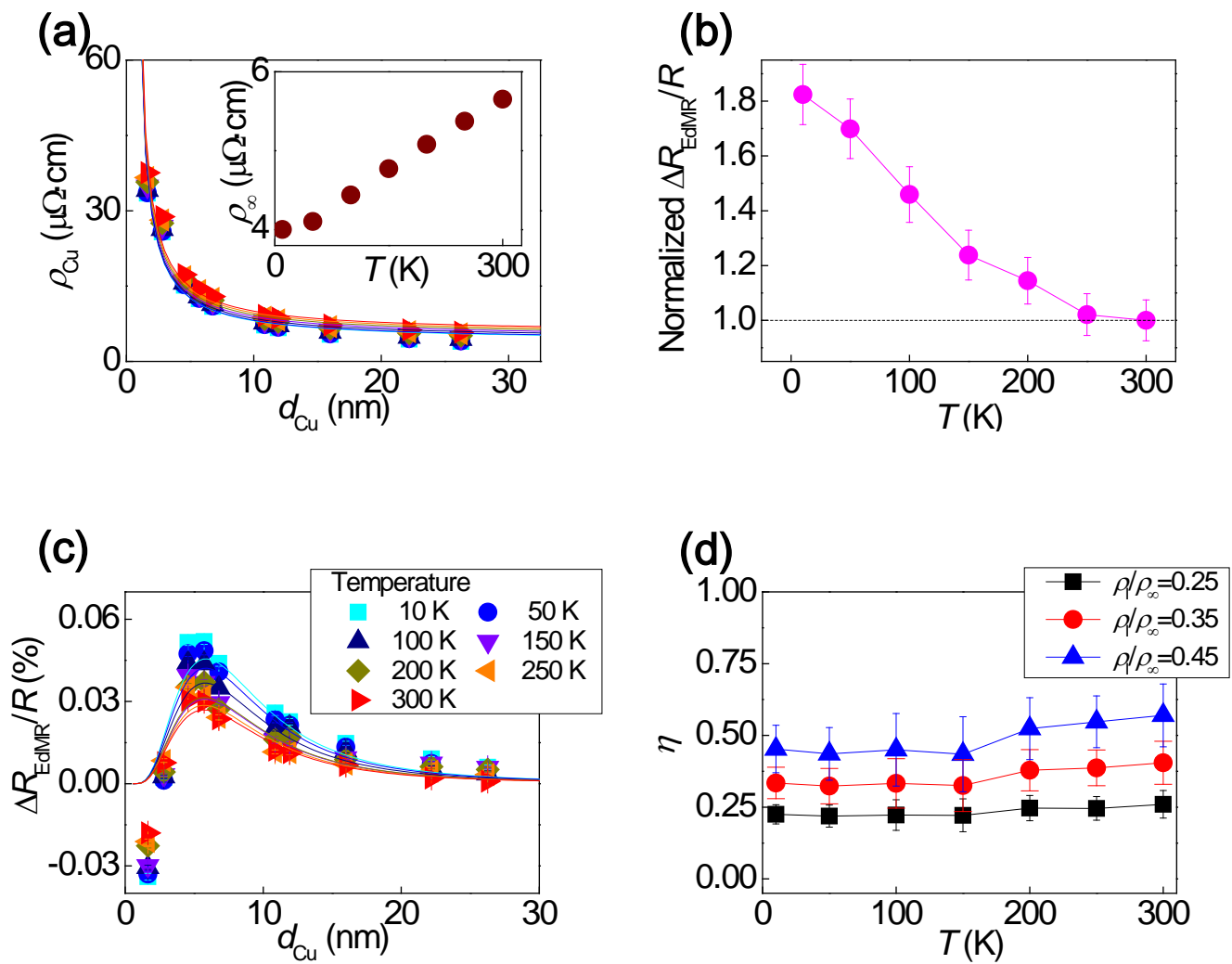

Fig. 5